\documentclass[aps,prl,twocolumn,showpacs,floatfix,superscriptaddress,longbibliography]{revtex4-1}
\usepackage{mathrsfs}
\usepackage{amssymb, amsbsy, amsmath, latexsym, dsfont, array, layout,mathrsfs,color,ulem,bm}
\usepackage[colorlinks,linkcolor=blue,anchorcolor=red,citecolor=blue]{hyperref}
\usepackage{multirow}
\usepackage{float}
\usepackage{threeparttable}
\usepackage{geometry}
\usepackage{ulem}
\newcommand{\ket}[1]{\left|{#1}\right\rangle}
\newcommand{\bra}[1]{\left\langle{#1}\right|}

\usepackage{graphicx}
\usepackage{xspace, stmaryrd, ulem}
\definecolor{delete}{rgb}{1.0, 0.0, 0.0}
\definecolor{edit}{rgb}{0.0, 0.0, 0.9}
\definecolor{comment}{rgb}{0.9, 0.0, 0.0}

\begin{document}

\title{Photonic chiral state transfer near the Liouvillian exceptional point}
\author{Huixia Gao}\thanks{These authors contributed equally to this work.}
\affiliation{Key Laboratory of Quantum Materials and Devices of Ministry of Education, School of Physics, Southeast University, Nanjing 211189, China}
\author{Konghao Sun}\thanks{These authors contributed equally to this work.}
\affiliation{CAS Key Laboratory of Quantum Information, University of Science and Technology of China, Hefei 230026, China}
\affiliation{CAS Center For Excellence in Quantum Information and Quantum Physics, Hefei 230026, China}
\author{Dengke Qu}\thanks{These authors contributed equally to this work.}
\affiliation{Beijing Computational Science Research Center, Beijing 100084, China}
\author{Kunkun Wang}
\affiliation{School of Physics and Optoelectronic Engineering, Anhui University, Hefei 230601, China}
\author{Lei Xiao}
\affiliation{Key Laboratory of Quantum Materials and Devices of Ministry of Education, School of Physics, Southeast University, Nanjing 211189, China}
\author{Wei Yi}\email{wyiz@ustc.edu.cn}
\affiliation{CAS Key Laboratory of Quantum Information, University of Science and Technology of China, Hefei 230026, China}
\affiliation{CAS Center For Excellence in Quantum Information and Quantum Physics, Hefei 230026, China}
\author{Peng Xue}\email{gnep.eux@gmail.com}
\affiliation{Key Laboratory of Quantum Materials and Devices of Ministry of Education, School of Physics, Southeast University, Nanjing 211189, China}

\begin{abstract}
As branch-point singularities of non-Hermitian matrices, the exceptional points (EPs) exhibit unique spectral topology and criticality, with intriguing dynamic consequences in non-Hermitian settings. In open quantum systems, EPs also emerge in the Liouvillian spectrum, but their dynamic impact often pertains to the transient dynamics and is challenging to demonstrate. Here, using the flexible control afforded by single-photon interferometry, we study the chiral state transfer when the Liouvillian EP is parametrically encircled. Reconstructing the density-matrix evolution by experimentally simulating the quantum Langevin equation, we show that the chirality of the dynamics is only present within an intermediate encircling timescale and dictated by the landscape of the Liouvillian spectrum near the EP. However, the chirality disappears at long times as the system always relaxes to the steady state. We then demonstrate the universal scaling of the chirality with respect to the encircling time. Our experiment confirms the transient nature of chiral state transfer near a Liouvillian EP in open quantum systems, while our scheme paves the way for simulating general open-system dynamics using single photons.
\end{abstract}

\maketitle

{\it Introduction.---}
Chiral state transfer~\cite{LKD95,LAS09,UMM11,BU11,MDH15,CHY17,ORN19,KSG21,NLL22,DGG23,SY23} is a signature dynamic phenomenon in non-Hermitian models with exceptional points (EPs)~\cite{MR08,MA19,IAS18}. Under a slow parameter change over a closed loop near an EP, the instantaneous state of the system is flipped but only one way around. The existence of such a chirality derives from the spectral topology of the complex eigenspectrum of the non-Hermitian Hamiltonian (or its Riemann surface), leading to adiabatic evolution along one encircling direction (over which the state is flipped), and amplification of nonadiabatic couplings in the other (the system turns back to its initial state)~\cite{MDH15}. The process has been observed in many open classical and quantum  systems~\cite{DMB16,ZWH18,SOP21,CAJ21,CAH22,DSW22,LWD21,RLZ22}, and considered useful for topological power~\cite{XMJ16,DMR16,LLW20,LDW20,YCH18} and state transfer~\cite{LWD21} as well as entanglement preparation~\cite{KCM24}. Focusing on the open quantum systems, while the full dynamics therein is described by the density-matrix dynamics governed by the Liouvillian~\cite{ML18,H19,MMC19,MMC20,AMM20,CAJ21,DCM92,PK98}, a no-jump condition is imposed through post selection, to ensure that the state evolution in the resulting conditional dynamics is driven by a non-Hermitian effective Hamiltonian~\cite{LHL19, NAJ19, ACN22}. Under realistic experimental conditions, such conditional dynamics are often imperfect, as quantum jumps inevitably occur as the system couples to the many degrees of freedom of the environment. These quantum jumps give rise to dephasing or decoherence, whose impact on the chiral state transfer has been demonstrated using superconducting qubits~\cite{CAJ21,CAH22}. Notably, while the transferred state typically becomes mixed and no longer an eigenstate of the non-Hermitian Hamiltonian, the chirality can be preserved under appropriate conditions~\cite{KSG21,SY23,RLZ22}.

A more systematic perspective can be obtained by viewing the dynamics using the full quantum master equation~\cite{BP07,HLM17,PKW21}, in which the Liouvillian superoperator is essentially non-Hermitian, capable of hosting EPs in its eigenspectrum. Dubbed Liouvillian EPs, they emerge as spectral degeneracies in the Liouvillian spectrum~\cite{CAJ21,CAH22,MMC20,KSG21}. Under what conditions can chiral state transfer also occur near the Liouvillian EPs is an important but open question, particularly as the Liouvillian EPs are usually located far away from the steady state. This limits their impact to transient dynamics, making a systematic experimental study challenging.

In this work, we experimentally illustrate the condition of chiral state transfer near the Liouvillian EP, by simulating the density-matrix evolution of a two-level open quantum system using single photons.
As a key ingredient of the experiment, we map the Liouvillian dynamics of the quantum master equation to the quantum Langevin equation~\cite{KG92}, and reconstruct the density-matrix evolution from the stochastic wave-function evolutions of single photons.
We then parametrically encircle the Liouvillian EP with different timescales, and demonstrate that
the chirality of the dynamics is present only at intermediate times. Importantly, the measured trajectories of the density-matrix evolution over the landscape of the Liouvillian spectrum exhibit similar key features to those over the Riemann surface of non-Hermitian Hamiltonians in a typical EP-encircling process. At long times, however, the system relaxes to a steady state regardless of the direction of parametric encircling, and the chirality disappears. We also show that the chirality universally scales with $1/T^\nu$, where $T$ is the encircling time, and $\nu$ is a parameters-dependent coefficient. Our experiment conclusively demonstrates the onset of chiral state transfer near a Liouvillian EP, confirms its transient nature, and sets the stage for future studies of the rich consequences of Liouvillian EPs in open quantum systems.

\begin{figure}[tbp]
	\includegraphics[width=0.5\textwidth]{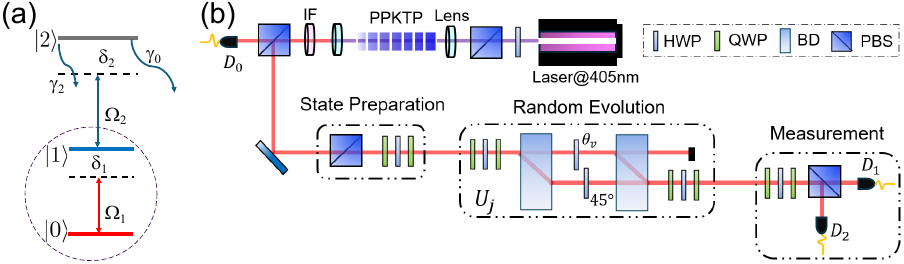}
	\caption{(a) Schematic illustration of the two-level open system considered in this paper. States $\ket{0}$ and $\ket{1}$ are linked by a coupling rate $\Omega_1$ with a detuning $\delta_1$.
States $\ket{1}$ is coupled to the environment through $\ket{2}$ (with coupling rate $\Omega_2$ and detuning $\delta_2$), which undergoes spontaneous emission either to the environment (and lost), or
back to state $\ket{1}$. The decay rates are $\gamma_0$ and $\gamma_2$, respectively.
(b) Experimental setup. Type-II spontaneous parametric down-conversion in a PPKTP crystal generates heralded single photons. Initial polarization states $\ket{x_-^{R}}$ are prepared when single photons pass through a polarizing beam splitter (PBS) and wave plates. The evolution $U_j$ is implemented by a Mach-Zehnder interferometer formed by a pair of beam displacers (BDs) and wave plates. Non-Hermiticity, corresponding to $\gamma_0$ in (a), is introduced by the polarization-dependent photon loss. The measurement is realized by wave plates and PBS. Finally, the photons are registered by avalanche photodiodes (APDs). The coincidence counts for D$_0$, D$_1$ and D$_0$, D$_2$ are then recorded, respectively.}
	\label{fig:setup}
\end{figure}

{\it Simulation of open systems.---}
We focus on an exemplary two-level open system, as illustrated in Fig.~\ref{fig:setup}(a). The ground state $|0\rangle$ is coherently coupled to the excited state $|1\rangle$, whereas the state $|1\rangle$ undergoes population loss to the environment through a light-assisted process mediated by an intermediate state $|2\rangle$. Further, we consider a spontaneous decay process from $|2\rangle$ to $|1\rangle$, which leads to dephasing of the two-level system $\{|0\rangle, |1\rangle\}$. The system is then described by the following hybrid Lindblad master equation
\begin{align}\label{master}
    \dot{\rho}   &=-i\left(H \rho-\rho H^{\dagger}\right)+L_\phi \rho L_\phi^{\dagger}-\frac{1}{2} L_\phi^{\dagger} L_\phi \rho-\frac{1}{2} \rho L_\phi^{\dagger} L_\phi,\nonumber\\
     &:=\mathcal{L}\rho,
   \end{align}
where $\mathcal{L}$ is a Liouvillian superoperator, and the non-Hermitian Hamiltonian is given by
\begin{align}
    H=H_0-i \Gamma|1\rangle\langle 1|.
    \label{nonH}
\end{align}
Here the effective population loss rate $\Gamma$ is dependent on the parameters $\gamma_{0,2}$, $\Omega_{1,2}$, and $\delta_{1,2}$ (see Fig.~\ref{fig:setup} and S2 of \cite{SM}). The Hermitian part of $H$ in the basis $\left\{|0\rangle,|1\rangle\right\}$ is
\begin{align}
    H_0=\begin{pmatrix}
\frac{\delta_1}{2} & -\Omega_1 \\
-\Omega_1^* & -\frac{\delta_1}{2}
\end{pmatrix}.
\end{align}
The dephasing channel is characterized by the quantum jump operator $L_\phi=\sqrt{\gamma_\phi}|1\rangle\langle 1|$ with the dephasing rate $\gamma_\phi=\frac{\gamma_2\gamma^2|\Omega_2|^2}{\Delta^4}$ and $\gamma=(\gamma_0+\gamma_2)/2$ (see S2 of~\cite{SM}).

An EP of the non-Hermitian Hamiltonian in Eq.~(\ref{nonH}) occurs at $\delta_1=0$ and $\Gamma=2|\Omega_1|$, where two branches of eigenvalues and eigenstates coalesce. Considering the Lindblad master equation in Eq.~(\ref{master}), the Liouvillian eigenspectrum $\lambda_i$ is calculated through $\mathcal{L}\rho^{R}_i=\lambda_i\rho^{R}_i$ with the corresponding Liouvillian eigenstates $\rho^{R}_i$. Similarly, EPs can also emerge in the Liouvillian eigenspectrum with coalescing eigenenergies and eigenstates, giving rise to unique behavior in the full quantum dynamics governed by the master equation. Here the long-time steady state of the Liouvillian corresponds to the eigenstate with the largest real part of the eigenvalue (nonzero for finite population loss rate $\Gamma$). The Liouvillian gap is identified by the difference between the largest and the second-largest real parts of the eigenenergies. We also note that, to simulate EP-encircling dynamics, the parameters of the Liouvillian (both those of the non-Hermitian Hamiltonian and the jump operators) are time-dependent.

\begin{figure*}[tbp]
\includegraphics[width=\textwidth]{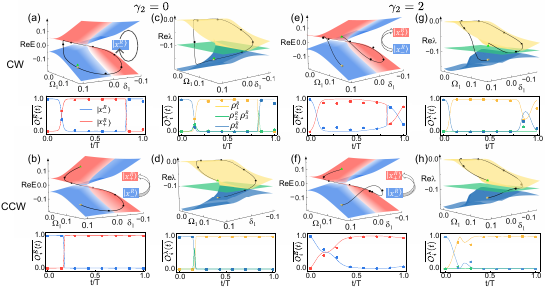}
\caption{Trajectories of the encircling dynamics for an encircling time $T=600$, for (a)(c)(e)(g) clockwise rotations and (b)(d)(f)(h) counterclockwise rotations. (a), (b) Trajectories against the eigenspectra of the non-Hermitian Hamiltonian $H$ with dephasing switched off ($\gamma_2=0$). The red (blue) color regions indicate the eigenenergy $E_+$ ($E_-$) with larger (smaller) imaginary part, corresponding to the right eigenstate $\ket{x_+^R}$ ($\ket{x_-^R}$) of $H$. The black solid trajectory corresponds to the ideal EP-encircling dynamics under $H$. The initial state $\ket{x_-^R}$ is represented by a yellow square, while the final state is represented by a green triangle. Lower row: the average overlap $\overline{O^E_{i}(t)}$ v.s. $t/T$. (c), (d) Trajectories against the eigenspectra of $\mathcal{L}$ with dephasing switch off ($\gamma_2=0$). The different color regions indicate the eigenenergy corresponding to the eigenmode $\rho_i^R$ ($i=1,2,3,4$) of $\mathcal{L}$.  (e), (f) Trajectories against the eigenspectra of $H$ for $\gamma_2=2$. (g), (h) Trajectories against the eigenspectra of $\mathcal{L}$ with $\gamma_2=2$. Experimental data are shown as hollow squares and solid dots, and theoretical predictions are represented by curves. The other parameters are $\gamma_0=20$, $\delta_2=0$, $\Omega_2=1$, and $\Delta t=1/1000$. Error bars indicate the statistical deviation, obtained by Monte Carlo simulations under the assumption of Poissonian photon-counting statistics. Some error bars are smaller than the size of the symbols.
}
\label{fig:data1}
\end{figure*}

To directly simulate the dynamics of open quantum systems described by the Lindblad master equations with photons is challenging. Instead, we simulate the dynamics depicted by a quantum Langevin equation in our experiment. This is because the dynamics of the open system depicted by the Lindblad master equation is equivalent to the stochastic wave-function evolution of the quantum Langevin equation~\cite{KG92} (see S1 of~\cite{SM})
\begin{align}
    i \frac{d}{d t}|\psi(t)\rangle&=\left(H-\frac{i}{2} L_\phi^{\dagger} L_\phi+i l(t) L_\phi\right)|\psi(t)\rangle\nonumber\\
    &:=\Tilde{H}(t)|\psi(t)\rangle,
\label{Htilde}
\end{align}
where $l(t)$ is the white noise, satisfying
$\langle l(t)\rangle=0$ and $\left\langle l(t) l^*\left(t^{\prime}\right)\right\rangle=\delta\left(t-t^{\prime}\right)$. Here $\langle\cdot\rangle$ denotes the ensemble average. For each realization of the noise $l(t)$, a distinct time-evolved state $|\psi_j\left(t\right)\rangle$
is obtained. The density matrix $\rho(t)$ of the system can be reconstructed by taking the ensemble average of these states $\rho\left(t\right)=\frac{1}{n}\sum_{j=1}^n |\psi_j\left(t\right)\rangle\langle\psi_j\left(t\right)| \equiv\sum_{i=j}^n \rho_j\left(t\right)$.
In our experiment, we take $n=10$, which, as we show below, provides a good enough estimation of
the density-matrix dynamics.

{\it Experimental simulation.---}
Figure~\ref{fig:setup}(b) displays the basic setup of the experimental simulation of the dynamics of the open system depicted by the Lindblad master equation in Eq.~(\ref{master}). The experiment comprises state preparation, evolution and measurement. As schematically illustrated in Fig.~\ref{fig:setup}(b), a photon pair is created via type-II spontaneous parametric down-conversion (SPDC), with one serving as a trigger and the other as a heralded photon. We then employ the horizontal and vertical polarization states of a heralded single photon as the basis states, i.e., $\{\ket{0}=\ket{H},\ket{1}=\ket{V}\}$. A polarizing beam splitter (PBS) and wave plates are used to prepare the initial state $\ket{x_-^R}$, where $\ket{x_-^R}$ is one of the right eigenstates of the non-Hermitian Hamiltonian $H$ in Eq.~(\ref{nonH}) with the smaller imaginary part of the eigenvalue, i.e., $H\ket{x_-^R}=E_-\ket{x_-^R}$.

Instead of implementing the dynamics of the open system depicted by the Lindblad master equation, we simulate the nonunitary dynamics driven by $\Tilde{H}(t)$ in Eq.~(\ref{Htilde}) up to a time $\tau$. Thus, we directly implement the time-evolution operator $U(\tau)$~\cite{XQW21} constructed stroboscopically according to
\begin{align}
	\label{eq.Ut}
	U(\tau)=\prod_{k=1}^{N}e^{-i\Tilde{H}(t_k)\delta t},
\end{align}
where $t_k=(k-1/2)\delta t$, $\delta t=\tau/N$, and we take $N=1000$ throughout the work. The nonunitary operator $U(\tau)$ is implemented on the basis of the following singular value decomposition
\begin{align} \label{nonunitary operator}
U(\tau)=
R(\beta_2,\theta_2,\beta'_2)L(\theta_v,\theta_h)R(\beta_1,\theta_1,\beta'_1),
\end{align}
where the rotation operator $R(\beta_i, \theta_i, \beta_i^{'})$ ($i=1,2$) can be implemented by a sandwich type of QWP($\beta_i$)-HWP($\theta_i$)-QWP($\beta_i'$). Here QWP and HWP are the abbreviations of quarter- and half-wave plates, respectively, and $\beta$ and $\theta$ are their setting angles. The polarization-dependent loss operator $L(\theta_v,\theta_h)$ is achieved with a Mach-Zehnder interferometer involving two beam displacers (BDs) and HWPs at
$\{\theta_v, \theta_h\}$. All the setting angles of wave plates are fixed according to the numerically calculated $U(\tau)$. 

For measurement, as illustrated in Fig.~\ref{fig:setup}(b), after the single photons have undergone
the series of gate operations that simulate the nonunitary evolution $U(\tau)$ in Eq.~(\ref{eq.Ut}), we construct the trajectory on the eigenspectral landscape of the Liouvillian superoperator $\mathcal{L}$ through the quantum state tomography. Specifically, the trajectory is defined according to
\begin{align}\label{lambdat}
\overline{\lambda}\left(t\right)&=\frac{\sum_{i=1}^4 \lambda_i(t)\left|\mathrm{Tr}\left(\rho^{L\dagger}_i\left(t\right)\rho\left(t\right)\right)\right|^2}{\sum_{i=1}^4\left| \mathrm{Tr}\left(\rho^{L\dagger}_i\left(t\right)\rho\left(t\right)\right)\right|^2}\nonumber\\
&:=\sum_{i=1}^4\lambda_i(t)\overline{O^{\lambda}_{i}\left(t\right)},
\end{align}
where $\rho^{L}_i$ satisfies $\mathcal{L}^\dag \rho^{L}_i=\lambda_i^* \rho^{L}_i$ ~\cite{SY23}.

For comparison, we also construct the trajectory of the state evolution over the Riemann surface of the non-Hermitian effective Hamiltonian $H$ in Eq.~(\ref{nonH}). The trajectory is defined as
\begin{align}\label{Et}
\overline{E}\left(t\right)&=\frac{\sum_{i=\pm}E_i\left(t\right)\langle x_i^L\left(t\right)|\rho\left(t\right)|x_i^L\left(t\right)\rangle }{\sum_{i=\pm}\langle x_i^L\left(t\right)|\rho\left(t\right)|x_i^L\left(t\right)\rangle}\nonumber\\
&:=\sum_{i=\pm}E_i\left(t\right) \overline{O^E_{i}\left(t\right)},
\end{align}
where $\ket{x_{\pm}^L}$ are the left eigenstates of the non-Hermitian Hamiltonian $H$ in Eq.~(\ref{nonH}), which are defined as $H^\dag|x_i^L\rangle=E_i^*|x_i^L\rangle$. The experimental results of $O^E_{i,j}\left(t\right)\equiv\left|\langle x_i^L\left(t\right)|\psi_j\left(t\right)\rangle\right|^2$ can be obtained by projective measurements. More specifically, we implement the projection $M_{i}=\ket{H}\bra{x_{i}^{L}(t)}+\ket{V}\bra{x_{i}^{L\perp}(t)}$ with wave plates. Here $\bra{x_i^{L\perp}(t)}$ ($i=\pm$) are the orthogonal states of $\bra{x_i^{L}(t)}$. A PBS is then used to map the basis states $\{\ket{H},\ket{V}\}$ to two distinct spatial modes, where photons are collected by two avalanche photodiodes (APDs) $D_1$ and $D_2$. The outputs are recorded in coincidence with trigger photons. Typical measurements yield a maximum of $150, 000$ photon counts per second (see S3 of~\cite{SM}).


{\it Experimental results.---}
We initialize the system in the eigenstate $\ket{x_-^R}$ of $H$ in Eq.~(\ref{nonH}), and impose the encircling path
\begin{align}
	\label{eq.path}
	\delta_1(t)&=0.09\sin(\pm2\pi t/ T +\pi/3),\\ \Omega_1(t)&=0.08+0.09\cos(\pm2\pi t/T+\pi/3),
\end{align}
where $+$ ($-$) indicates the clockwise (counterclockwise) encircling. We then consider two cases for studies: (i) with a very long total encircling time $T=600$; (ii) with an intermediate encircling time $T=90$.

For case (i), we perform a standard encircling of the EP of the non-Hermitian Hamiltonian $H$ in Eq.~(\ref{nonH}), by turning off dephasing ($\gamma_2=0$). As illustrated in Figs.~\ref{fig:data1}(a) and (b), two types of the trajectories of the encircling dynamics in the adiabatic limit are shown, along clockwise and counterclockwise rotations, respectively. We construct the trajectory (black curve) against the Riemann surface of the non-Hermitian Hamiltonian. The red (blue) color regions indicate the eigenenergy $E_+$ ($E_-$) with larger (smaller) imaginary part, corresponding to the eigenstate $\ket{x_+^R}$ ($\ket{x_-^R}$). The experimental results demonstrate that, starting from the initial state $\ket{x_-^R}$, the system returns to its initial state for a clockwise encircling, while it flips to another eigenstate $\ket{x_+^R}$ for a counterclockwise encircling.
This indicates that the encircling dynamics driven by the non-Hermitian Hamiltonian $H$ with dephasing switched off exhibit the chiral state transfer.

In Figs.~\ref{fig:data1}(c) and (d), we show the corresponding trajectory over the landscape of the Liouvillian eigenspectrum. The chiral state transfer is also observed here, which is a direct result of the vanishing of the Liouvillian gap for the case of $\gamma_2=0$. Specifically, in Fig.~\ref{fig:data1}(c), the final state is not the steady state of the Liouvillian. It is understandable as the relaxation toward the steady state takes a much longer time (than $T=600$ here), now that the Liouvillian gap vanishes.

Then, we switch on dephasing ($\gamma_2\neq 0$) and construct the trajectories over the Riemann surface [Fig.~\ref{fig:data1}(e)(f)] and the Liouvillian spectrum [Fig.~\ref{fig:data1}(g)(h)], respectively. As illustrated in Fig.~\ref{fig:data1}(e)(f), the chirality disappears as the final states of clockwise and counterclockwise encirclings are close to each other, and eventually merge in the long-encircling-time limit (for example $T=600$). This behavior can be easily interpreted by examining the trajectories on the eigenspectral landscape of the Liouvillian shown in Fig.~\ref{fig:data1}(g)(h), the state turns back to the close adjacency of the initial state in both clockwise and counterclockwise directions. This is because, for $\gamma_2\neq 0$, the Liouvillian spectrum acquires a gap, meaning the steady states are now approached exponentially fast. Hence, given long enough encircling time (which is the case here), the system always relaxes to its instantaneous steady state, leading to the disappearance of chirality.

\begin{figure}[tbp]
	\includegraphics[width=0.5\textwidth]{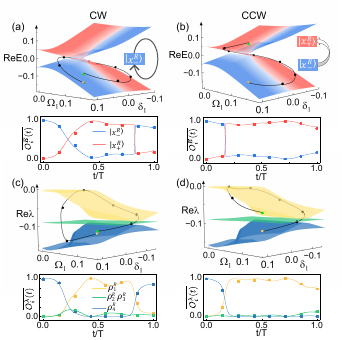}
	\caption{Trajectories of the encircling dynamics for an encircling time $T=90$, for (a)(c) clockwise rotations and (b)(d) counterclockwise rotations. (a), (b) Trajectories against the eigenspectra of $H$ under dephasing $\gamma_2=2$. (c), (d) Trajectories against the eigenspectra of $\mathcal{L}$ with $\gamma_2=2$. The experimental results are obtained by $10$ quantum Langevin equation realizations.  The initial state and the other parameters are the same as Fig.~\ref{fig:data1}.
}
	\label{fig:data2}
\end{figure}

To recover chirality in the presence of dephasing, we consider case (ii), where the total encircling time is appropriately short, as illustrated in Fig.~\ref{fig:data2}. Here an intermediate encircling time $T=90$ and a finite dephasing rate $\gamma_2=2$ are chosen. The observed chiral state transfer is purely due to the encircling of the Liouvillian EP, which is evident when comparing the trajectories of Fig.~\ref{fig:data2}(c)(d) and Fig.~\ref{fig:data1}(a)(b)(c)(d).
The trajectory nearly adiabatically follows the Liouvillian spectral landscape along one direction [Fig.~\ref{fig:data2}(d)], but involves a non-adiabatic jump (fast relaxation to the steady state) in the alternative direction.
Since the Liouvillian EP lies away from the steady state, the chiral state transfer manifests only in the transient dynamics. This is also the reason why the adiabaticity in Fig.~\ref{fig:data2}(d) is only approximate. Finally, we note that, different from the chiral state transfer in the purely non-Hermitian case, under dephasing the final state is mixed.

\begin{figure}[tbp]
	\includegraphics[width=0.5\textwidth]{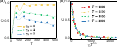}
	\caption{(a) Chirality as a function of the total encircling time $T$ and the dephasing rate $\gamma_2$. (b) Scaling relation of the chirality with respect to $\gamma_2$ and $T$. The initial state is $\rho(0)=\ket{1}\bra{1}$, the encircling path is $\delta_1(t)=0.09\sin(\pm2\pi t/ T)$ and $\Omega_1(t)=0.08+0.09\cos(\pm2\pi t/T)$. 
	The experimental results are obtained by $10$ quantum Langevin equation realizations.
}
	\label{fig:data4}
\end{figure}

{\it Chirality.---}
To quantify our observations, we define chirality $C$ as follows~\cite{CAH22}
\begin{align}
		\label{eq.C}
C=\frac{1}{2}\text{Tr}[\sqrt{(\tilde{\rho}_{cw}-\tilde{\rho}_{ccw})^\dagger(\tilde{\rho}_{cw}-\tilde{\rho}_{ccw})}],
	\end{align}
where $\tilde{\rho}_{cw,ccw}=\rho_{cw,ccw}/\text{Tr}(\rho_{cw,ccw})$, with $\rho_{cw,ccw}$ being the final-time density matrix of the clockwise (cw) and counterclockwise (ccw) encircling.
To measure the chirality experimentally, we reconstruct the density matrices $\rho_j(T)$ for the $j$-th evolution with single-qubit quantum state tomography, and obtain the ensemble-averaged density matrix
\begin{align}
\label{eq.rhotc}
\rho(T)=\frac{1}{n}\sum_{j=1}^{n}{(\text{max}}|\xi_j|)\rho_j(T),
\end{align}
where $\xi$ is the eigenvalue of $U(\tau)U^\dagger(\tau)$~\cite{H50,GWX24,SLM18} (see S3 of~\cite{SM}). Substituting $\rho(T)$
into Eq.~(\ref{eq.C}), the chirality $C$ for different $\gamma_2$ and $T$ can be determined.

In Fig.~\ref{fig:data4} we illustrate the chirality as a function of the encircling time $T$ and dephasing rate $\gamma_2$. When the dephasing is turned off $(\gamma_2 = 0)$, in the long-time limit, $C$ tends to be a finite value close to unity. On the contrary, with non-zero $\gamma_2$, $C$ inevitably approaches zero if the encircling time is sufficiently long. As illustrated in Fig.~\ref{fig:data4}(a), a chirality peak emerges at intermediate $T$, which corresponds to the optimal parameter regime for chiral state transfer. Remarkably, the chirality at the final time exhibits a universal scaling with respect to $\gamma_2$ and $T$, i.e., $C=f(\gamma_2T^{1/\nu})$, as illustrated in Fig.~\ref{fig:data4}(b). We obtain $\nu=1.7221$ by fitting the experimental data, which agrees with the numerical calculated result. We note that, while the form of the scaling relation is universal, the coefficient $\nu$ is path-independent but parameter-dependent.

{\it Conclusion.---}
We experimentally study chiral state transfer near a Liouvillian EP. Different from the well-known chiral state transfer in purely non-Hermitian models, in a genuine open quantum system, chiral state transfer generally occurs in the transient dynamics, since the Liouvillian EP lies away from the steady state. An exception has recently been reported in bistable Rydberg vapors~\cite{XSW24}, where the non-linearity induces an exceptional structure in the steady-state landscape of the Liouvillian eigenspectrum. Therein, the chiral state transfer occurs at long times. As such, the dynamic consequences of the Liouvillian EP and exceptional structures in quantum open system have rich implications for future studies. In this regard, our experimental scheme offers a general framework wherein dynamics of quantum open-systems can be simulated and studied in a controlled fashion using single photons.

\begin{acknowledgments}
This work is supported by the National Key R\&D Program of China (Grant No. 2023YFA1406701) and the National Natural Science Foundation of China (Grants No. 92265209, No. 12025401, No. 12374479).
\end{acknowledgments}

Data Availability Statement: The data supporting this study's findings are available within the Letter~\cite{data}.

\end{document}